%% file: wini00.tex
\documentstyle[epsfig]{europhys}
\input euromacr

\begin{document}
\euro{?}{?}{1-$\infty$}{2000}
\Date{? ? 2000}
\shorttitle{T. Winiecki {\it et al.}, Motion in a quantum fluid}
\title{Motion of an object through a dilute quantum fluid}
\author{T. Winiecki \And C. S. Adams}
\institute{Department of Physics, University of Durham, Rochester Building,
South Road, Durham, DH1 3LE, England. UK.}
\rec{}{}
\pacs{
\Pacs{03}{75.F}{Bose-Einstein condensation}
\Pacs{47}{37.+q}{Superfluid hydrodynamics}
\Pacs{67}{40.Vs}{Vorticity in $^4$He}}
\maketitle

\begin{abstract}

We simulate the motion of a massive object through a dilute Bose-Einstein condensate
by numerical solution of the non-linear Schr\"odinger equation coupled to an
equation of motion for the object. Under a constant applied force, the 
object accelerates up to a maximum velocity where a vortex ring is 
formed which slows the object down. 
If the applied force is less than a critical value, the object becomes
trapped within the vortex core. We show that the motion can be 
described using the time-independent quantum states, and use these states
to predict the conditions required for vortex scattering. 

\end{abstract}

\section{Introduction}

One of the most elementary questions that can be asked about a fluid,
is how will an object move through it? In quantum fluids, it is expected that 
the object moves without resistance at velocities, $v$, up 
to a critical value, $v_{\rm c}$, where energy and momentum conservation allow excitations. 
If the object mass is large, the critical velocity is given by the Landau criterion, 
$v_{\rm c}=(\epsilon/p)_{\rm min}$, where $\epsilon$ and $p$ are the energy and momentum 
of the excitation \cite{till90}. Experiments on the motion of objects in liquid helium
suggest that the excitations are often vortices \cite{donn91}, 
however, understanding the exact mechanism of vortex formation is
impeded by the lack of a complete hydrodynamical model.
In contrast, for dilute quantum fluids such as the recently discovered
atomic vapour Bose-Einstein condensates \cite{fermi}, the dynamics 
can be accurately described by the Gross-Pitaevskii
equation, a form of non-linear Schr\"odinger equation (NLSE) \cite{dalf99}. 
Consequently, this system provides a near ideal
testing ground for advancing our knowledge of superfluid flow.
Experimental measurements of the heating produced by laser beam moving 
through an atomic condensate suggest that the dominant 
mechanism of momentum transfer is vortex shedding \cite{rama99}, in
agreement with NLSE simulations of a two-dimensional homogeneous flow past a fixed
object \cite{fris92,wini99a,wini00}.
 
However, if the object has a finite mass, the fluid back-action is significant,
and completely different dynamics can arise.
For example, ions in liquid helium nucleate vortex rings and
become trapped within the vortex core \cite{donn91,ray64}.
Here, we study the general case of the motion of an object with finite
mass moving through a dilute Bose-Einstein condensate. 
The time-evolution is found by solving a NLSE coupled to an equation of
motion for the object. We show that the object can be accelerated up to a
maximum velocity where a vortex ring emerges encircling the object. 
If the applied force is not too large,
the object subsequently becomes trapped within the core of the ring. 
We show that the motion can be predicted from the 
time-independent uniform flow states and 
apply this method to give a complete description of 
energy and momentum conservation during vortex scattering.

\section{Numerical method}

Throughout the paper we use dimensionless units, where distance 
and velocity are measured in terms of the fluid healing length, $\xi$
and the speed of sound, $c$, respectively.
The dynamics of the fluid are described by a wavefunction, 
$\psi(\mbox{\boldmath$r$},t)$, whose time-evolution in the fluid rest frame 
is given by
\begin{equation}
{\rm i} \partial_t \psi(\mbox{\boldmath$r$},t) = 
- \textstyle{1\over2} \nabla^2 \psi(\mbox{\boldmath$r$},t) 
+ V(\mbox{\boldmath$r$}-\mbox{\boldmath$v$}t)\psi(\mbox{\boldmath$r$},t) 
 + \vert\psi(\mbox{\boldmath$r$},t) \vert^2 \psi(\mbox{\boldmath$r$},t)~.
\end{equation}
where $V$ is the object potential.  
The position of the object is given by
\begin{equation}
M\dot{\mbox{\boldmath$v$}}=\mbox{\boldmath$F$}
+\int {\rm d}^3r~\frac{{\rm d}V}{{\rm d}\mbox{\boldmath$r$}}
\vert\psi(\mbox{\boldmath$r$},t) \vert^2~,
\end{equation}
where $\mbox{\boldmath$F$}$ is an external force and the second term is the force on the 
object due to the fluid. The computation is simplified by transforming into the frame
of the object, where Eq.~(1) may be rewritten as
\begin{equation}
{\rm i} \partial_t \tilde{\psi}(\mbox{\boldmath$r'$},t)  = 
- \textstyle{1\over2} \nabla^2 \tilde{\psi}(\mbox{\boldmath$r'$},t) 
+ V(\mbox{\boldmath$r'$})\tilde{\psi}(\mbox{\boldmath$r'$},t) 
 + \vert\tilde{\psi}(\mbox{\boldmath$r'$},t) \vert^2 \tilde{\psi}(\mbox{\boldmath$r'$},t)
+{\rm i}\mbox{\boldmath$v$}\nabla'\tilde{\psi}(\mbox{\boldmath$r'$},t)~,
\label{eq:3}
\end{equation}
where $\tilde{\psi}(\mbox{\boldmath$r'$},t)=
\psi(\mbox{\boldmath$r$},t)$ is the wavefunction in 
the fluid frame written in terms of the object frame coordinates,
$\mbox{\boldmath$r'$}=\mbox{\boldmath$r$}-\mbox{\boldmath$v$}t$.
The system is prepared in a time-independent laminar flow state, 
$\tilde{\psi}(\mbox{\boldmath$r'$},t)=\phi(\mbox{\boldmath$r'$}){\rm e}^{{\rm i}\mu t}$,
where $\mu$ is the chemical potential, by solving 
Eq.~(\ref{eq:3}) using Newton's method \cite{huep97,wini99b}. 
From this initial state, the time evolution due to an applied force, $\mbox{\boldmath$F$}$,
is evaluated by integrating Eq.~(\ref{eq:3}) using a semi-implicit
Crank-Nicholson formula. The conservation of momentum
\begin{equation}
\mbox{\boldmath$P$}_0+\mbox{\boldmath$F$}t=M\mbox{\boldmath$v$}+\frac{\rm i}{2}
\int {\rm d}^3r'\left[\psi\nabla'\psi^*-\psi^*\nabla'\psi\right]~,
\end{equation}
where $\mbox{\boldmath$P$}_0$ is the initial momentum of the fluid, 
is imposed as an additional constraint. The equations are discretized on a three
dimensional grid using the non-linear transform $\hat{x}=x/(D+\vert x\vert)$,
where $D$ is a scaling parameter, to map
an infinite box onto the space $-1\rightarrow 1$.
The grid contains 140 points in each dimension and we use a time step ${\rm d}t=0.02$.
For the object we choose a penetrable sphere with mass $M=200$ and radius $R=3.3$ 
($V=1.0$ for $\vert\mbox{\boldmath$r'$}\vert\leq 3.3$ and $0$ elsewhere).
For an atomic condensate this would correspond to a small cluster of impurity 
atoms in a sphere with a diameter of a few microns. However,
the main conclusions are not overly sensitive to these parameters.

\section{Results}

\begin{figure}[hbt]
\centering
\epsfig{file=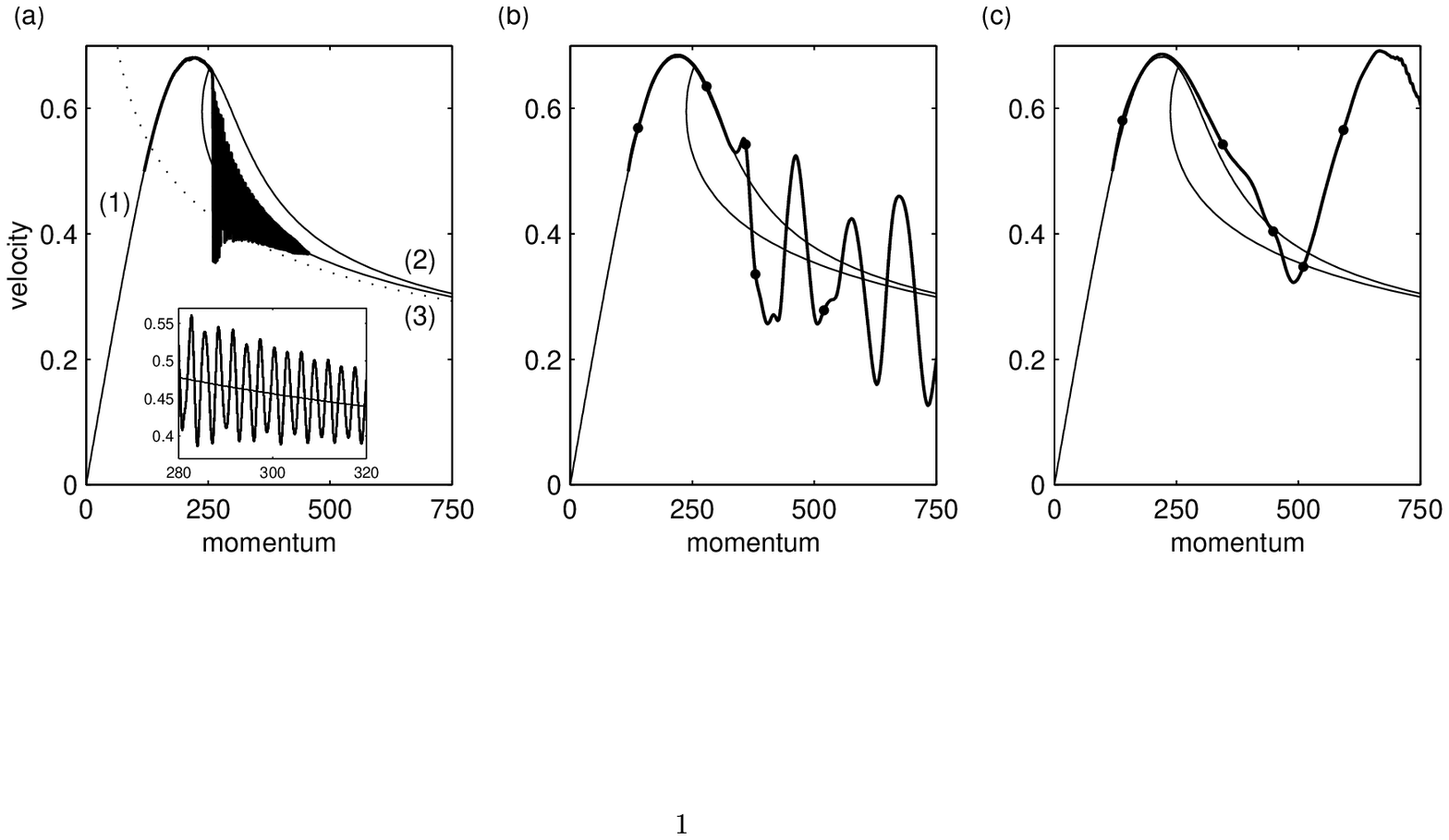,clip=,width=14cm,bbllx=70,bblly=190,bburx=575,bbury=390}
\caption{The evolution of the object velocity due to a constant applied 
force. The velocity is plotted (bold line) against the total momentum $\mbox{\boldmath$P$}=\mbox{\boldmath$F$}t$ for (a) $\mbox{\boldmath$F$}=0.05$,
(b) $\mbox{\boldmath$F$}=2$ and (c) $\mbox{\boldmath$F$}=4$. 
Also shown are the time-independent solutions of
the uniform flow equation, Eq.~(\ref{eq:3}), corresponding to (1) laminar flow,
(2) an encircling vortex ring, and (3) a pinned ring. 
As the object accelerates momentum is transferred from the object
to the fluid. At the peak velocity, $v_{\rm c}=0.68$, an encircling vortex ring
emerges and begins to slow the object down. 
An abrupt decrease in velocity occurs when the object
jumps into the vortex core. This jump excites vibrations of the ring
leading to large oscillations of the object velocity (inset). The oscillations
are damped by the emission of sound waves. In (c), the force is
sufficient to detach the object from the ring and the cycle repeats.
}
\label{fig:1}
\end{figure}

\begin{figure}[hbt]
\centering
\epsfig{file=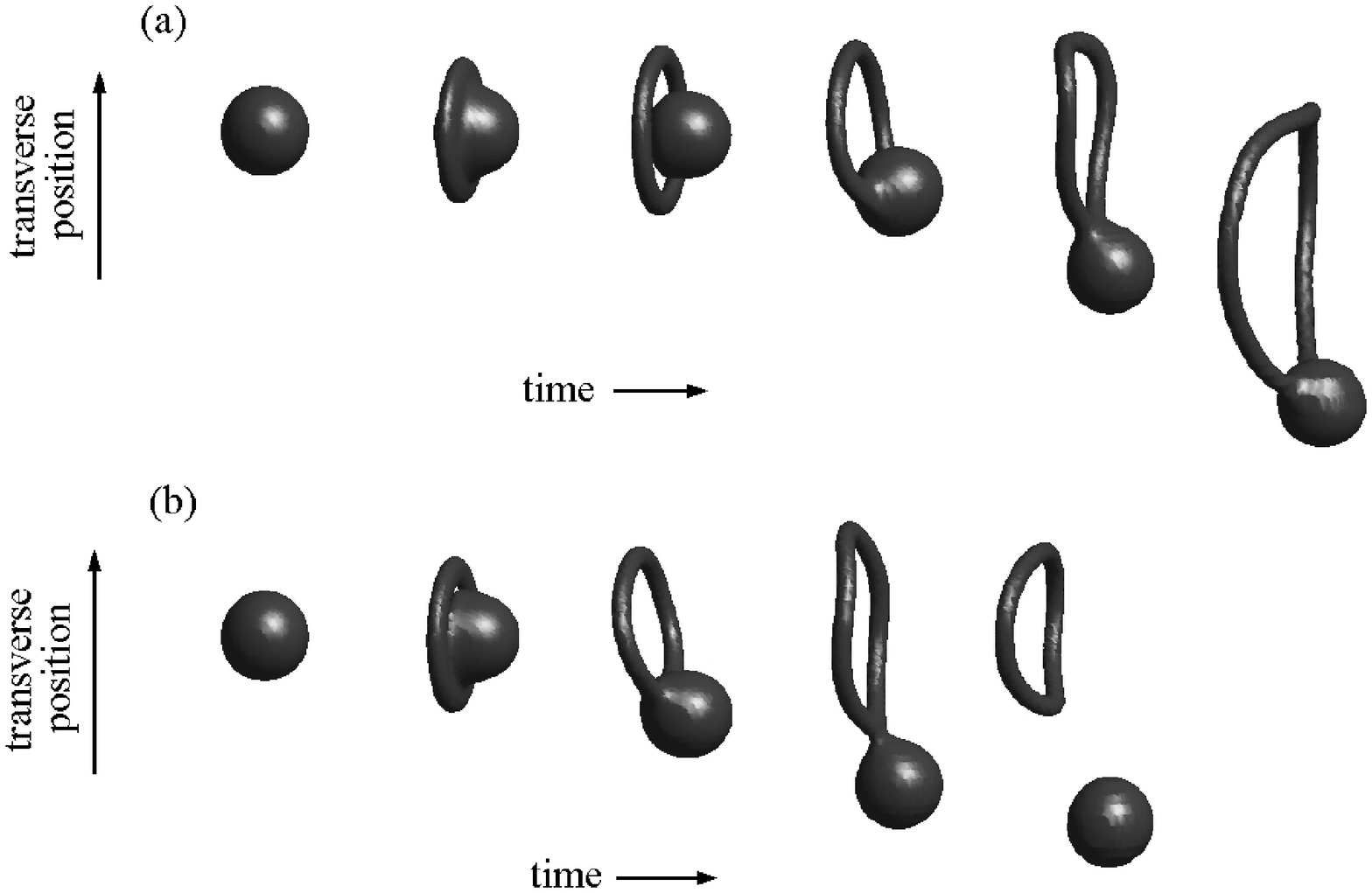,clip=,angle=-90,width=12cm,bbllx=100,bblly=130,bburx=490,bbury=710}
\caption{Sequence of surface contour plots of the fluid density for 
(a) $\mbox{\boldmath$F$}=2$ and (b) $\mbox{\boldmath$F$}=4$. 
The motion is from left to right, and the real space deflection due to the attraction
of the vortex core is indicated by the $y$ coordinate. The momentum (or time) of
each frame is indicated by a dot in Fig.~\ref{fig:1}(b) and (c), except 
for the last frame in (a), where $P=1562$. 
Note that after detachment, (b), the ring size remains constant and the object
and ring move at different velocities.
}
\label{fig:2}
\end{figure}

The evolution of the object velocity, $v$, due to a constant applied 
force, $F$, is shown in Fig.~\ref{fig:1}. As the velocity increases,
momentum is transferred from the object to the fluid. This momentum
transfer can be described in terms of an increase in the 
effective or hydrodynamical mass of the 
object, $m_{\rm eff}=(\partial P/\partial v)^{-1}$, where $P=Ft$. 
The effective mass becomes infinite at the critical velocity, $v_{\rm c}=0.68$,
and then negative as the object begins to slow down. Further 
insight is provided by the 
surface contour plots of the fluid density shown in Fig.~\ref{fig:2}. At the peak
velocity, $v_{\rm c}$, a vortex ring emerges encircling the object.
The object is weakly bound in the direction of
motion, therefore as the ring grows, the object and the ring decelerate.

When the vortex core begins to separate from the object boundary, 
the encircling ring configuration becomes 
unstable with respect to transverse motion, and stochastic
fluctuations induce a transition to a pinned ring, where the object is
bound within the vortex core (see Fig.~\ref{fig:2}).
In our simulations, defining the external force, $\mbox{\boldmath$F$}$, 
at a slight angle to the numerical
grid axis is sufficient to induce the transition.
On moving into the core, the object acquires a transverse
velocity thereby deflecting its trajectory (Fig.~\ref{fig:2}). 
The deflection angle is a few degrees so this effect could be observable.
If the ring detaches, a second ring forms and the object is pulled
back in the opposite direction. Consequently vortices are emitted on
alternating sides of the object, similar to the vortex shedding behaviour
observed in classical fluids.

The jump into the core also leads to the excitation of 
oscillatory modes of the vortex ring  Fig.~\ref{fig:2}(a).
One mode of oscillation dominates \cite{sam91} and the frequency is 
independent of the applied force. As the fluid is 
compressible, an accelerating object creates sound waves which
damp the motion. This damping is apparent in the oscillations of the object velocity 
in Fig.~\ref{fig:1}(a) inset. If the applied force is maintained the vortex radius continues to increase
and eventually the motion becomes indistinguishable from that
of a free vortex ring, indicated by the dotted line in Fig.~\ref{fig:1}(a).
Also shown in Fig.~\ref{fig:1} are velocity-momentum curves 
predicted by the time-independent solutions of the NLSE for a uniform flow,
Eq.~(\ref{eq:3}).
The three branches correspond to (1) laminar flow,  (2) an encircling vortex ring
and (3) a pinned vortex ring \cite{wini99b}.
Excluding the ring excitations, the motion closely follows
the time-independent states. It follows that appropriate uniform flow 
solutions may be used to predict the motion of more complicated
objects. To test whether a spherical object favours the encircling vortex ring
configuration, we performed calculations on a sphere ($R=3.3$) with a hemispherical 
surface bump ($R=1.5$). The largest effect occurs when the bump lies
in the equatorial plane. In this case, the critical velocity is 
reduced from 0.68 to 0.65, and the vortex ring
emerges asymmetrically with its axis pulled towards the bump.
However, the initial ring radius is still similar to the no bump case.
Subsequently, the object or ring rotate such that the
vortex core is pinned to the bump.

\section{Vortex detachment}
 
\begin{figure}[hbt]
\centering
\epsfig{file=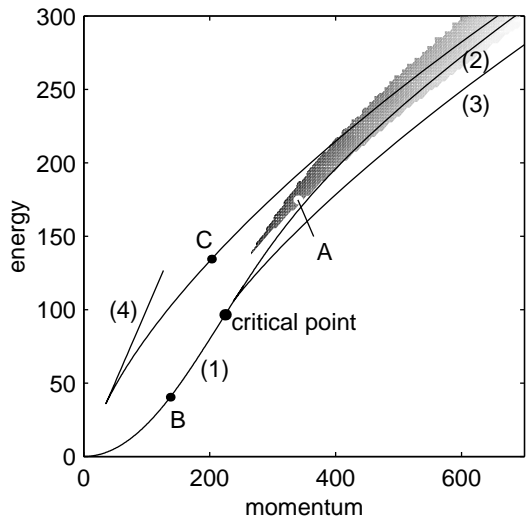,clip=,width=7cm,bbllx=15,bblly=180,bburx=215,bbury=380}
\caption{Energy and momentum conservation during 
vortex scattering. The allowed energy and momentum states of 
a moving object with mass $M=200$ (1-3) and a free vortex ring (4)
determined from the time-independent solutions of the uniform flow equation, Eq.~(\ref{eq:3}).
Under a constant force, the object moves up the
laminar flow branch of the dispersion curve (1) passing through the critical point 
where a vortex ring is formed, then continues along
or just above the encircling ring branch (2). When 
the system reaches the shaded region, it is possible
for the bound object-vortex ring (A) to decay into an object dressed by a 
laminar flow (B) and a free vortex ring (C). If, before detachment, the object moves into the 
core, the energy tends towards the pinned ring branch (3) making
detachment less probable. The grey scale indicates
the radius of the scattered vortex ring (darker is smaller).
The velocity is determined by the slope of the dispersion curves. 
}
\label{fig:3}
\end{figure}

If the applied force is larger, the object can 
detach from the vortex ring as in Fig~\ref{fig:2}(b). 
After detachment the size of the ring remains constant, and
the object can accelerate again up to the critical
velocity where another ring forms and so on. 
The beginning of this repetitive cycle is apparent in 
the time evolution shown in Fig.~\ref{fig:1}(c).
If detachment occurs, the initial encircling or pinned ring
system evolves into an object and a free vortex
ring which moves more slowly as in Fig.~\ref{fig:2}(b). The nucleated ring can be
considered as an excitation of the fluid with 
energy $\epsilon$ and momentum $p$. The dispersion curve for vortex rings
in the NLSE was determined by Jones and Roberts \cite{jon82} and is reproduced in 
Fig.~3. The simple form of the Landau criterion, 
$v_{\rm c}=(\epsilon/p)_{\rm min}$, cannot be applied to unbounded 
fluid as the minimum occurs in the limit $p\rightarrow\infty$.
A complete description of energy and momentum conservation during 
vortex scattering should include the recoil energy $p^2/2M$, and the energy and 
momentum stored in the laminar flow around the object. Such a description is 
provided by considering the time-independent quantum states.
In this picture, the process of vortex scattering appears as a `decay'
of an encircling ring state into a free vortex ring and a 
laminar flow state. In Fig.~\ref{fig:3}, we plot the dispersion curves
for a moving object and a free vortex ring obtained from the time-independent
solutions \cite{wini99b}. The shaded region indicates the range of 
initial energies and momenta where
vortex scattering may occur, i.e., for which there exists final vortex ring 
and laminar states which satisfy the conservation laws. In a time-dependent
simulation with a constant external force,
the object moves up the dispersion curve passing through the critical point
where a vortex is formed and then continues along the encircling ring
branch (2). When the system enters the shaded region, 
(point A in Fig.~\ref{fig:3}), energy and momentum conservation permit
decay into an object dressed by laminar flow (B) and a free vortex ring (C). 
Note that the energy at point A is higher than the time-independent value because the object is dragged out of the plane of the ring (see Fig~\ref{fig:2}). 
If the applied force is low, the object moves into the vortex
core and the system relaxes towards the lower
branch of the dispersion curve (3) making detachment
unlikely.  

\section{Discussion}

It is interesting to note that the velocity profile shown
in Fig.~\ref{fig:1}(a) is quite similar to those observed for
ions in superfluid helium (see e.g. \cite{rayf67}).
To convert between dimensionless units and values for
helium, we use the measured values of the number density,
$n_0=2.18\times10^{28}$~m$^{-3}$, the quantum of circulation, 
$\kappa=h/m=9.98\times10^{-8}~{\rm m}^2{\rm s}^{-1}$,
and the healing length $\xi/\sqrt{2}=0.128$~nm \cite{ray64}
leading to a mass unit, $\tilde{m}=mn_0\xi^3=0.13m$, 
where $m$ is the mass of a helium atom. Our object parameters, $R=3.3$, $M=200$,
correspond to 25 helium atoms with radius 0.6~nm,
similar to the `snowball' that surrounds a positive ion \cite{donn91}.
In the NLSE model, the speed of sound is given by $c=\hbar/m\xi$, 
therefore using the healing length as the fit parameter, one obtains 
$v_{\rm c}\sim 60$~ms$^{-1}$, in rough agreement with experiments \cite{rayf67}.
Although the NLSE cannot be expected to give a complete description of
superfluid $^4$He, it could be considered 
to compliment existing theories of vortex formation by ions \cite{muir84},
because the vortex core structure is included explicitly.
 
\section{Summary}

In summary, we have studied the motion of an object through a dilute quantum 
fluid. We show that there is a continuous transition between laminar
flow and an encircling ring followed by a jump where
the object moves into the vortex core. This jump leads to a deflection
of the object trajectory and excitations of the vortex ring.
If the object has a surface bump near the equator, the encircling vortex 
emerges asymmetrically. If the applied force is large, the object evades capture by the ring
leading to periodic vortex shedding. The simulations reproduce the velocity
evolution observed for ions in superfluid helium,
suggesting that the behaviour may be characteristic of
more complex quantum fluids. We show that the motion and 
the conditions required for vortex scattering can 
be predicted using the time-independent states for 
uniform flow. This approach could be extended to provide useful insight into
other complex problems in quantum fluid mechanics such as vortex reconnections
and sound emission.

\stars We thank D. C. Samuels, C. F. Barenghi, B. Jackson and J. F. McCann for 
stimulating discussions. Financial support 
for this work is provided by the Engineering and Physical
Sciences Research Council (EPSRC).

\end{document}

%% file: euromacr.tex
\def\And{{\rm and\ }}

\def\stars{\bigskip\centerline{***}\medskip}

\newif\ifboo \boofalse


%% file: wini00.bbl
\begin{thebibliography}{99}

\bibitem{till90}
D. R. Tilley and J. Tilley, {\it Superfluidity and superconductivity},
3rd Ed., (IoP, Bristol, 1990).

\bibitem{donn91} R. J. Donnelly, 
{\it Quantized vortices in Helium II}, (CUP, Cambridge, 1991).

\bibitem{fermi} See {\it Bose-Einstein condensation in atomic gases}, Proc. Int. School
of Physics Enrico Fermi, eds. M. Inguscio, S. Stringari and C. Wieman (IOS Press,
Amsterdam, 1999).

\bibitem{dalf99} F. Dalfovo, S. Giorgini, L. P. Pitaevskii, and S. Stringari,
Rev. Mod. Phys. {\bf 71}, 463-512 (1999).

\bibitem{rama99} C. Raman, M. K\"ohl, R. Onofrio, D. S. Durfee, C. E. Kuklewicz, 
Z. Hadzibabic, and W. Ketterle, 
Phys. Rev. Lett. {\bf 83}, 2502-2505 (1999).

\bibitem{fris92} T. Frisch, Y. Pomeau, and S. Rica, 
Phys. Rev. Lett. {\bf 69}, 1644-1647 (1992).

\bibitem{wini99a}T. Winiecki, J. F. McCann, and C. S. Adams,  
Phys. Rev. Lett. {\bf 82}, 5186-5189 (1999).

\bibitem{wini00}T. Winiecki, B. Jackson, J. F. McCann, and C. S. Adams,  
J. Phys. B. {\bf 33}, to appear (2000).

\bibitem{ray64} G. W. Rayfield and F. Reif, 
Phys. Rev. {\bf 136}, 1194-1208 (1964).

\bibitem{huep97} 
C. Huepe and M.-\'E. Brachet, 
C. R. Acad. Sci. Paris, {\bf 325}, S\'erie 2 b, 195-202 (1997).

\bibitem{wini99b} T. Winiecki, J. F. McCann, and C. S. Adams,
Europhys. Lett. {\bf 48}, 475-481 (1999).

\bibitem{sam91}D. C. Samuels and R. J. Donnelly,
Phys. Rev. Lett. {\bf 67}, 2505-2508 (1991).  

\bibitem{jon82} C. A. Jones and P. H. Roberts, 
J. Phys. A {\bf 15}, 2599-2619 (1982).

\bibitem{rayf67} G. W. Rayfield, 
Phys. Rev. Lett. {\bf 19}, 1371-1373 (1967).

\bibitem{muir84} C. M. Muirhead, W. F. Vinen, and R. J. Donnelly, 
Phil. Trans. R. Soc. Lond. A {\bf 311}, 433-467 (1984).

\end{thebibliography}
